\documentclass[review]{elsarticle}

\usepackage{hyperref}

\journal{Journal of \LaTeX\ Templates}

\begin{document}

\begin{frontmatter}

\title{What T$_c$ Tells}

\author{Theodore H Geballe}
\author{Robert H Hammond}
\author{Phillip M Wu}
\address{Department of Applied Physics and Materials Science, Stanford University, Stanford CA 94305}





\begin{abstract}

Superconductivity has continued to be a fascinating phenomenon ever since its discovery in 1911. The magnitude of the transition temperature, T$_c$, provides valuable insight into the underlying physics. Here we provide select examples of the extensive research that has been done towards understanding T$_c$, and some cases where further investigation is called for. We believe that searching for new and enhanced T$_c$'s remains a fertile frontier.

\end{abstract}

\begin{keyword}
Superconductivity \sep cuprates \sep negative-U \sep Fe based superconductors
\end{keyword}

\end{frontmatter}


\section{Historical Introduction}

The transition temperature, T$_c$, of a superconductor emerges from all the underlying microscopic electronic interactions within it. Consequently, T$_c$ itself and its response to controlled changes in external parameters often help reveal the responsible mechanisms. This first occurred in 1911 when Kammerlingh Onnes observed a mysterious discontinuous drop in the resistance of mercury to an immeasurably low value \cite{Onnes1911}. Onnes had been motivated to use highly purified mercury, to test two rival theories. One had predicted an approach to infinite conductivity and the other to infinite resistivity when a good metal is cooled in liquid helium \cite{Rogalla2011}. Our objective in this chapter is to review other examples where unexpected T$_c$s have sometimes revealed new physics, sometimes hidden materials science, and sometimes poor experiments. The changes in the T$_c$ in response to controlled parameters, such as change in mass, composition, pressure or structure, become important inputs to theory and help to reveal new insights into superconducting mechanisms.

The scarcity of liquid helium in the early 20th century meant there were few places suited to search for and study T$_c$. Nowadays, modern cryostats and equipment make it easy to perform resistance or SQUID magnetization measurements, arguably the first tests to look for T$_c$. However the ease with which such measurements can be done is a two-edged sword, causing many false positives to be reported. Not surprisingly this has caused reports of unexpectedly high T$_c$s to be greeted with a healthy dose of skepticism particularly when they have not been duplicated elsewhere.

The greatest impact of an unexpected T$_c$ after Onnes was the discovery 75 years later by Bednorz and Mueller of superconductivity in the ternary ceramic oxide Ba$_2$CuO$_4$ family. They cautiously reported the disappearance of resistance with an onset above 30 K, higher than any known superconductor at the time, as "Possible superconductivity..." \cite{Bednorz1986}. Their discovery was almost immediately verified in Japan \cite{Uchida1987} and soon throughout the world. T$_c$ was soon raised to the technologically important and the scientifically challenging temperature range above liquid nitrogen by Wu and Chu \cite{Wu1987}. The vast amount of new physics that followed is discussed briefly later, in other chapters of this volume and in 416,000 other works cited by Google (given by a search for "cuprate superconductivity")!

The nature of T$_c$ in the cuprates is still not fully understood, and much work remains to be done. Here we review reports of unexpected T$_c$s in highly oxidized cuprates and in other materials worthy of further investigation. Finally there are reports of superconductivity that have not been verified and have come to be known as unidentified superconducting objects, "USOs" \cite{Kopelevich2015}. A number of these reports are due to faulty or misinterpreted experiments, but in others further investigations are warranted. We believe it is worthwhile to take USO reports seriously, because as the miners said back in the California gold rush days, "There may be gold in them thar hills", or as is said in a similar vein today, "DonÕt throw the baby out with the bathwater."

\section{Discoveries before BCS}

Corresponding sudden drops in the resistance of mercury were soon found in Leiden in Sn, Pb and other "pure" elements\footnote{The drop was described in the Leiden Communciations as \emph{supra-}conductivity meaning beyond conductivity which is a more apt description of the phenomenon than \emph{super-}.}. Little attention seems to have been given to the clue that mercury amalgams, solder and other poor metals, were found to have T$_c$s in the same range as the pure metals. We are unaware of what motivated Walther Meissner, when liquefied helium was so precious, to search for superconductivity in intermetallic compounds. In our opinion his discovery of superconductivity in CuS made by reacting non-superconducting Cu with sulfur \cite{Meissner1929} to give a poorly conducting metal, and also to discover superconductivity in a number of other intermetallic compounds \cite{Meissner1932, Roberts1976} is just as important as the discovery of perfect diamagnetism for which he is justly famous \cite{Meissner1933}. It seems reasonable to assume that these unusual T$_c$s led Enrico Fermi, six years before the discovery of the BCS theory, to encourage Bernd Matthias and John Hulm, then at the University of Chicago to search for more superconductors \cite{Matthias}, hoping that a pattern of occurrence might provide important clues of the responsible mechanisms.

\section{Matthias ÒRulesÓ}

\begin{figure}
\includegraphics[width=1\textwidth]{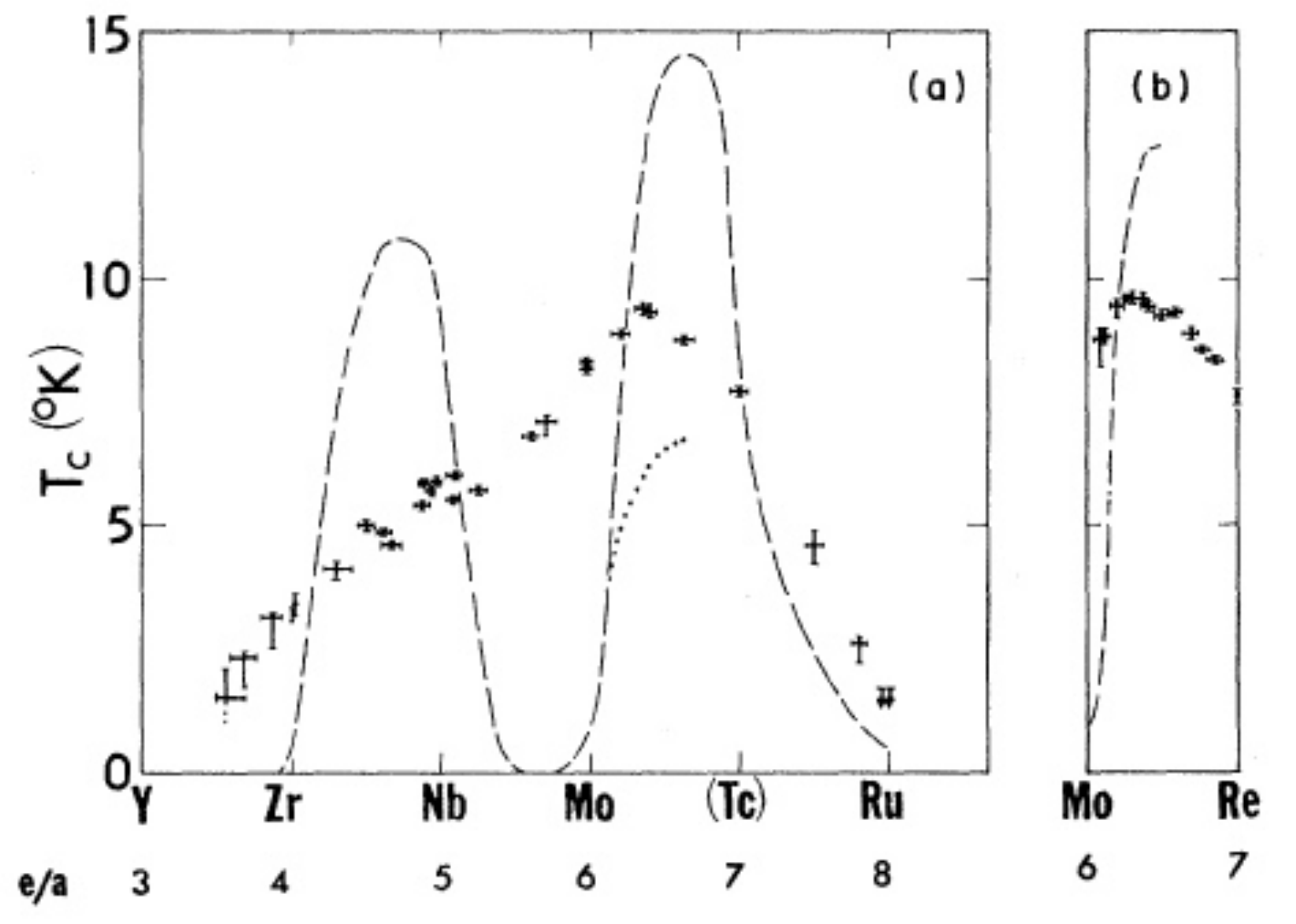}
\caption{\label{fig:collhamm} Superconducting transition temperatures of transition metal alloys as a function of the filling of the d-band. The dashed curves represent crystalline T$_c$ as a function of e/a or filling of the d-band showing maxima at 4.75 and 6.9. The points represent the amorphous counterparts.\cite{Matthias1955,Collver1973}. Figure reprinted with permission from \cite{Collver1973}.}

\end{figure}

Investigations chiefly by Matthias and Hulm's groups in the US and by Alekseevski's in Russia were responsible for changing superconductivity from being rare to being a common ground state of non-magnetic metals \cite{Matthias1963}. Bernd Matthias proposed empirical rules known by his name \cite{Matthias1955} that are useful mainly because they are so simple and in many cases have predictive power. The most useful rule, which applies to transition metal alloys and compounds with unfilled d bands, is that T$_c$ simply depends on the average number of electrons per atom (e/a), as shown in Fig. \ref{fig:collhamm} \cite{Matthias1955}. It is not obvious why there should be a connection between Mendeleev's periodic table and superconductivity; the periodic table is based upon chemistry and its organization depends upon energy scales typically three orders of magnitude greater than superconducting energy scales. But that didn't bother Matthias, who had a keen empirical sense that allowed him to organize the T$_c$s of 3d, 4d, and 5d elements and alloys in a systematic way. The rule can be understood by making the crude assumption that T$_c$ is a function of the filling of rigid d-bands. This of course is very rough and conflicts with Anderson's theory of dirty superconductors which assumes scattering by non-magnetic centers averages over the Fermi surface \cite{Anderson1959}, and with Collver and Hammond's demonstration that the e/a rule is not applicable for amorphous metals because the bands are not rigid. As shown by the dotted curve in Fig. \ref{fig:collhamm} for amorphous metals there is only a single maximum with an interesting small peak at half filling that may be due to some short range structure but has yet to be investigated \cite{Collver1973}. Exceptions to the Matthias e/a rule, as we shall see later, provided clues of unexpected physics and materials science.

However the e/a rule successfully predicted, for example, the relative increase in T$_c$ when a small concentration of an element to the right of Ti is alloyed with Ti. The T$_c$ increase is proportional to the number of added electrons. For example, the T$_c$ increase when 1\% of Mo in column VI of the periodic table is alloyed with Ti is twice as great as the T$_c$ increase when 1\% of Nb in column V is alloyed with Ti. This rule was found to be approximately true for a very large number of alloys except for Fe, where the increase was an order of magnitude larger than expected. This anomalous result was at first thought to be smoking gun evidence for a new type of pairing mechanism until careful annealing experiments \cite{Raub1966} showed it to have a straightforward materials science explanation, namely an inhomogeneous distribution of Fe. The Fe nucleated small regions of bcc Ti in which the Fe was concentrated in the correct amount to be consistent with the e/a rule. The lesson here is that apparently enhanced superconductivity can be caused by an unrecognized inhomogeneity.  

\subsection{Type II Superconductivity.} 

The first practical use of the data was to construct a magnet made of ductile MoRe wire of composition corresponding to the second peak in Fig. \ref{fig:collhamm}. Measurements of the current and field characteristics of a short length of MoRe reliably predicted the magnetic field generated by a coil wound from the wire. Soon in response to the need for shielding of the maser amplifiers planned for transcontinental microwave communication link being designed at Bell Labs, the T$_c$ of the newly discovered compound Nb$_3$Sn was found to remain remarkably high when short lengths were subjected to large fields and currents \cite{Kunzler1961}. This surprising result was not expected from the-then-accepted Mendelsohn ÒspongeÓ theory that explained why so called ÔhardÕ transition metal superconductors remained superconducting in high fields but had very low critical currents.\footnote{They were hard and brittle because of impurities, usually oxygen that were not detected. Tungsten which contained undetected oxygen was thought to be a polymorph of the element until it was found to have the same A15 structure as Nb$_3$Sn, explaining why Nb$_3$Sn is referred to in the literature as being a $\beta$-Tungsten compound.} It was assumed that the superconductor existed as a sponge of very thin filaments with dimensions $<<$ penetration length and incapable of carrying large currents. The correct type II theory had already been theoretically discovered by Abrikosov \cite{Abrikosov1957} but had been unappreciated until it was found to explain why T$_c$ in Nb$_3$Sn remained high in high fields and large currents. Soon a wealth of new physics, materials science, and technologies emerged.\footnote{It is most unfortunate that the earlier investigations of L. V. Shubnikov, who had clearly discovered type II superconductivity in experiments done in Kharkov much earlier were not recognized. He was put to death as an enemy of the people under the Stalin regime before he received any credit.\cite{Sharma2006}}

\section{The BCS Era} 

The square root dependence of T$_c$ upon isotopic mass, discovered by E. Maxwell and by Reynolds et al. after enriched isotopes of non-transition metal elements such as tin and lead became available after WWII \cite{Maxwell1950,Reynolds1950}, provided the essential clue that led Bardeen, Cooper, and Schrieffer to construct the long sought for microscopic theory of superconductivity, the BCS theory. When the electrons and phonons are weakly coupled, the BCS equation is simply 

$kT_c = 1.14 \hbar\omega exp(-1/N_0V)$

where the mass dependence comes from the pre-factor, and the exponent is a dimensionless measure of the pairing potential ($N_0$ is the density of states at the Fermi surface, V is the interaction potential). While the theory is motivated by measurements of the isotope mass, we note that the BCS pairing interaction could be more general and not strictly limited to electron-phonon coupling.\footnote{The idea that lattice vibrations could be important was not new. In 1922 Onnes and Tuyn reported in a Leiden Communication "the object was to trace a possible difference between the vanishing point of lead and uranium lead. It seemed not impossible that the occurrence of superconductivity might be influenced by the mass of the nucleus." We estimate that the $\sim$.01 K difference in T$_c$ between normal and 206 Pb was not quite enough to distinguish with the gas thermometer they used. It also should be noted that Fr{\"o}hlich \cite{Frohlich1950} in his theory of superconductivity found a square root mass dependence of T$_c$.}

\section{T$_c$s of Elements}

When T$_c$ of Ru was found to be independent of isotopic mass \cite{Geballe1961} it was for a very short time thought to be evidence for a non-phonon pairing mechanism. However Anderson and Morel \cite{Morel1962} soon showed the result followed from BCS if retardation is included, resulting in a modified formula,

$kT_c \sim \hbar\omega exp(-\frac{1}{\lambda - \mu^*})$, 

where $\mu^*=\mu/(1+\mu ln(2E_F/\hbar \omega))$ is the Coulomb repulsion $\mu$ between the paired electrons that is renormalized by a mass dependent term that for Ru roughly cancels the pre-exponential term. The Coulomb repulsion between the paired electrons is reduced because of the large difference in electron and phonon velocities. One electron of the pair induces a lattice response of positive charge and is far removed when the other electron senses the positive charge: the pairing interaction is localized in space and retarded in time \cite{Morel1962}. Localization in space offers a rational for explaining why Nb and niobium compounds often have higher T$_c$s than related superconductors \cite{Matthias1963}.

The absence of superconductivity in Mo \cite{Strongin1960} was difficult to understand in terms of BCS, because extrapolating T$_c$s from dilute solutions of elements adjacent in the periodic table suggested otherwise. However the samples used were high purity commercial Mo that contained $\sim$0.01\% Fe. When the Fe was removed T$_c$ was found to be 0.9 K \cite{Geballe1962}. Evidently, the Fe is a very strong pair breaker in Mo, suppressing T$_c$ by $\sim$100 K per percent Fe. The isotope effect in pure Mo was found to have an intermediate value of $\sim$0.3, quite understandable in terms of the Morel-Anderson theory \cite{Matthias1963}.

\begin{figure}
\includegraphics[width=1\textwidth]{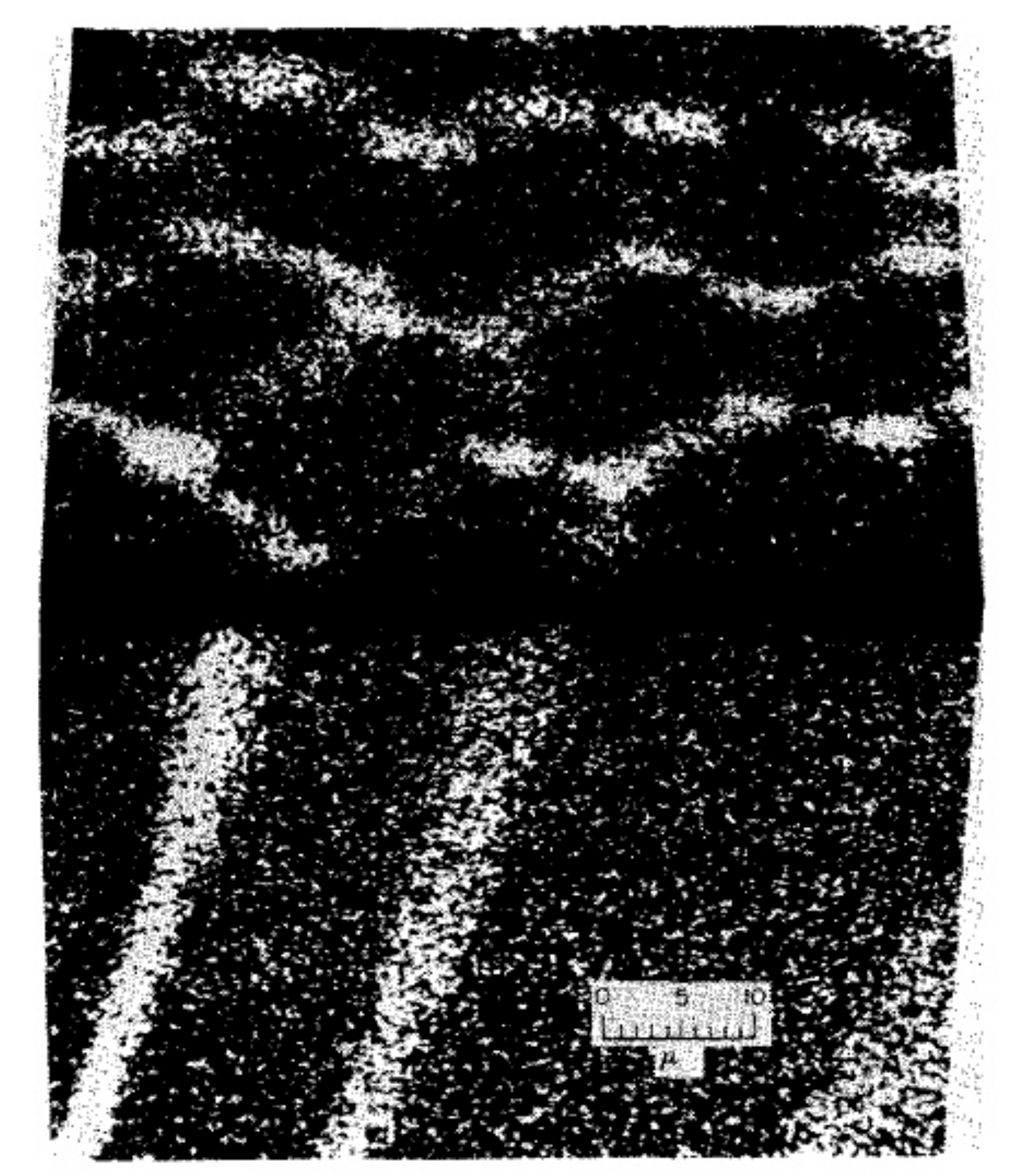}
\caption{\label{fig:larh2} Three-dimensional perspective of LaL$\alpha$1 x-ray emission from a prismatic section through the sample of La$_{0.005}$Rh$_{9.995}$ excited by 23 kV electrons, and showing honeycomb structure of the lanthanum-containing phase in the bulk of rhodium. The upper part of the image shows the horizontal surface, the dark horizontal zone is the edge and the lower part of the image is from the adjacent vertical surface. Since the electron beam, used for scanning the surfaces, has an approximate cross section of 1 micron, the image does not give geometrical information on details smaller than a few
microns, and the actual structure of the tubular walls is consequently not resolved. Nonetheless, La is concentrated into the walls of continuous vertical tubes, with mostly hexagonal cross sections. Figure and text reprinted with permission from \cite{Arrhenius1964}.}
\end{figure}

A search for superconductivity in rhodium uncovered clear evidence for proximity-induced superconductivity in bulk material\footnote{The Leiden secondary thermometers were made from a special batch of free machining Cu i.e., copper containing a small amount of Pb, and were most likely temperature and field dependent proximity induced thermometers}. An investigation of the La-Rh phase diagram to determine the ground state of Rh, showed that as little as 1\% La gave a sharp T$_c$ transition that excluded flux in a zero field cooled experiment (that is the field was applied after the sample was cooled below T$_c$. Arrhenius et al. showed by beautiful TEM investigations \cite{Arrhenius1964} that this was due to a jungle gym of superconducting LaRh$_5$ filaments, Fig. \ref{fig:larh2}. Upon further dilution to 0.1\% La the filaments were no longer connected as can be seen in the TEM and the transitions were broadened. This was convincingly shown to be due to proximity induced superconductivity in the Rh phase by additions of small amounts of Fe. Fe is non-magnetic in LaRh$_5$ and had no observable effect on T$_c$ of the 1\% sample, but forms a localized magnetic state in Rh and destroyed the superconductivity in the 0.1\% sample, giving the first clear proof, so far as we know, of proximity coupling in 3 dimensions. The T$_c$ of pure Rh has subsequently been found to be 325 $\mu$K \cite{Buchal1983}.

\subsection{Dense and Collapsed Phases}

It has been known since the early days of superconductivity that the so-called white tin is 6-fold coordinated and is metallic and superconducting whereas the less dense gray tin polymorph with the more open diamond 4-fold coordination is not \cite{Roberts1976}. This pattern of superconductivity is commonly found when more highly coordinated structures are obtained by quenching from high pressures or from the liquid or vapor phase in post transition metal elements such as Sb, Si and Ge. Greatly enhanced T$_c$s are also found in open structured Be and Bi, when their coordination is increased \cite{Roberts1976}, and illustrates the ubiquitous nature of metallic superconductivity.

The element Lithium in a compressed phase under high pressure has T$_c$ reaching 23 K at 80 GPa \cite{Shimizu2002,Struzhkin2002} and this presumably is related to its light mass. Hydrogen is predicted to have much higher T$_c$s at much higher pressures \cite{Buzea2005}. Shock tube experiments indicate transient metallic behavior. More recently, Ca was observed to superconduct with T$_c$ exceeding 25 K at 160 GPa, making it the highest known T$_c$ of the elements to date \cite{Yabuuchi2006}. More in depth discussion of metallic elements \cite{Hamlin2015}, insulating elements \cite{Shimizu2015}, and hydrogen-rich materials \cite{Struzhkin2015} under pressure are found later in this volume.

\section{Other mechanisms: Negative-U Pairing}

Various other mechanisms besides electron-phonon coupling have been considered to give rise to superconductivity. In this section, we specifically focus on one idea: that of negative-U pairing. 

When IV-VI semiconducting PbTe is doped with between 0.3 to 1.5 \%Tl it becomes superconducting whereas when doped with the same concentrations of other elements it remains a dirty semiconductor. \cite{Hulm1970,Matsushita2005,Nemov1998} This unique behavior is related to the valence skipping property of the Tl Ion.  In the gas phase and in solids such as TlSe, Tl$^{+2}$ disproportionates, 2Tl$^{+2} \rightarrow$ Tl$^{+1}$+Tl$^{+3}$ driven by the increased stabilities of the filled and empty 6S configurations with respect to the 6S$^1$ one. Please see the review of superconductivity in doped semiconductors by Bustarret in this volume \cite{Bustarret2015}.

\begin{figure}
\includegraphics[width=1\textwidth]{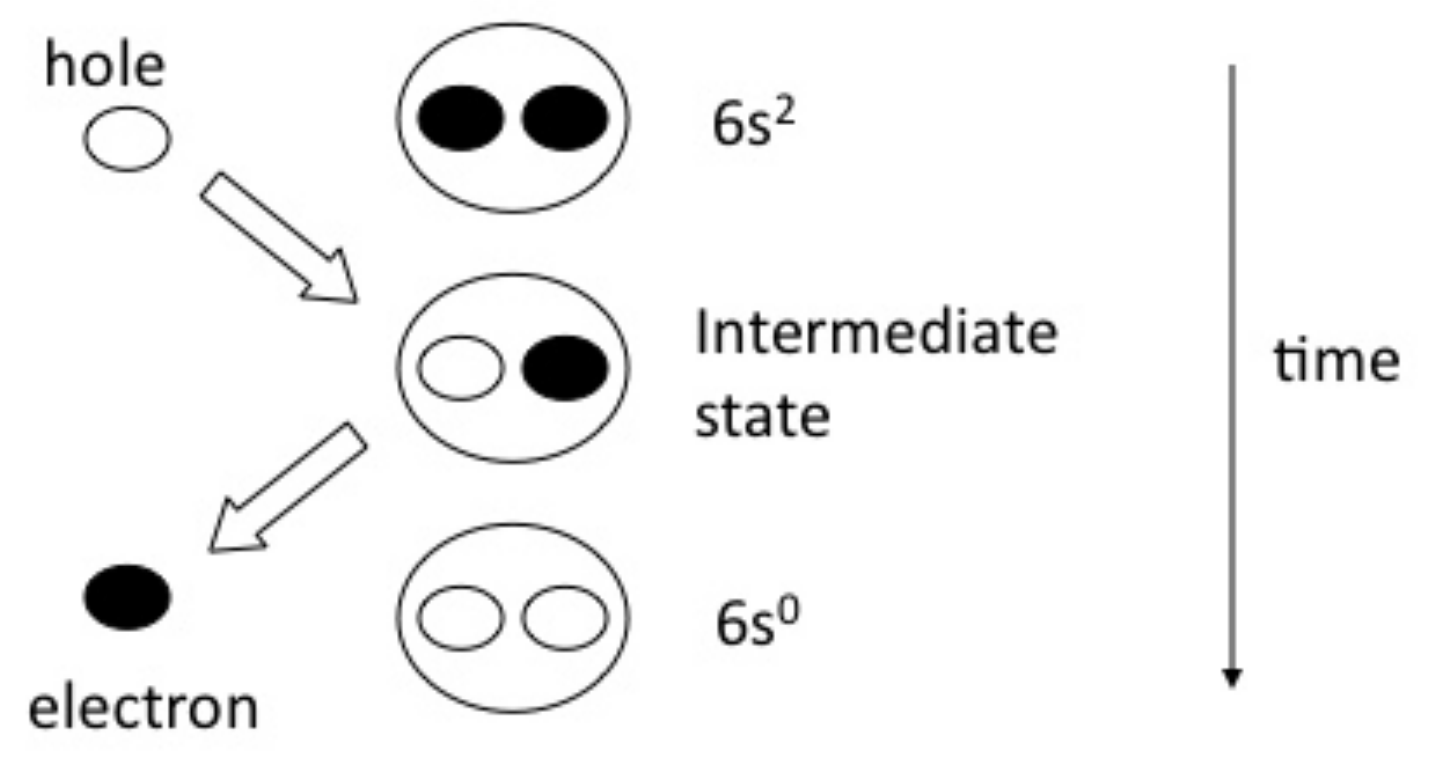}
\caption{\label{fig:chargekondo} Schematic representation of the superconducting pairing induced by charge Kondo fluctuations in Tl-doped PbTe. The charge Kondo intermediate state fluctuates between 6S$^0$ and 6S$^2$, providing an electronic means for superconductivity to arise.}
\end{figure}

Models suggesting these two states could be paired states responsible for the superconductivity \cite{Moizhes1981} were discounted because of the large reduction in Franck-Condon matrix elements expected from the large difference in radii. \cite{Anderson} However the minima in the resistivity versus temperature data were subsequently shown to be the result of charge Kondo scattering \cite{Dzero2005}, an electronic process not dependent upon the lattice response, as shown schematically in Fig. \ref{fig:chargekondo} and there is no Franck-Condon suppression. The negative U pairing mechanism requires i) that the two states be degenerate, ii) that they have the same coordination, i.e. no Franck-Condon suppression and iii) that they are virtually bound, i.e. hybridized with the conduction band. Nature is kind as this occurs automatically in PbTe as Tl is substituted; Tl$^{+1}$ is stable initially and the non-bonding extra electron fills states in the valence band until at 0.3\% doping the +3 and +1 states become degenerate. \cite{Dzero2005}

Experimentally negative U pairing can be more effective than the phonon induced pairing - the T$_c$s of the of Tl doped PbTe crystals are more than an an order of magnitude higher than comparably doped semiconductor superconductors \cite{Hulm1970}. The negative U pairing frequently results in charge density ground states, such as in TlSe, AgO and many other binary and ternary compounds \cite{Robin1967} that suppresses the superconducting state. Interfacing these localized states with good metals has been theoretically shown to lead to a very high T$_c$ \cite{Berg2008}. However attempts to increase the density of negative U centers further so far has led to localized charge density waves. Further work favoring delocalization by screening layers (see discussion on Hg1201 below) are needed.

Oxygen vacancies may serve as negative U centers in a variety of transition-metal oxides, as an example, T$_c$s of SrTiO$_{3-x}$ are found at carrier concentrations as low as $5*10^{17} /$cm$^3$, an order of magnitude lower than for any other cation dopants in SrTiO$_3$. This attests to the effectiveness of the pairing mechanism \cite{Lin2013}, suggesting oxygen vacancies must be considered as candidates for explaining enhanced T$_c$s in SrTiO$_3$ and in other systems such as the serious USOÕs discussed below.

The perovskite BaKBiO$_3$ and BaPbBiO$_3$ families in which Bi ions also can form degenerate 6S$^0$ and 6S$^2$ configurations \cite{Sleight1975,Uchida1987a,Cava1988} offer further evidence for possible high T$_c$ when negative U pairing may be involved. Recent studies suggest the situation is more complex, with polymorphs of BaPbBiO$_3$, oxygen breathing modes, and disorder playing a significant role in determining T$_c$ \cite{Giraldo-Gallo2012,Yin2013,Luna2014}. These studies suggest that the superconductivity in the bismuthates arises from large electron-phonon interaction, or alternatively, the negative U centers have dynamics that must be accounted for. For more on the bismuthates, we refer the reader to Sleight's chapter in this volume \cite{Sleight2015}.


\section{Cuprates}

Major research in cuprates has been focused on hole doped ones where T$_c$ as a function of concentration forms a "dome," beginning in an under doped region $\sim$0.05 holes per CuO$_2$ layer reaching a maximum at optimal doping, and dropping to zero in an overdoped region with $\sim$0.3 holes per layer, see \cite{Tallon1995}. T$_c$ at optimal doping is an important issue that remains poorly understood. For instance, the optimal T$_c$ of La$_{2-x}$Sr$_x$CuO$_4$ (LSCO) is 40 K, whereas that of Hg-1201 compound which differs from LSCO by having an additional layer of HgO$_{1-x}$ inserted between the LaSrO layers is more than twice as large. It is difficult to ignore this clue.  One approach relates changes in T$_c$ to differences in copper-apical oxygen bond lengths, which in turn induce subtle changes in the structure of the Fermi surface \cite{Kuroki2011,Sakakibara2012}. However this model is inadequate to explain the small apical oxygen distances found in highly oxidized cuprates discussed below. A promising explanation is that highly polarizable HgO layers screen the long range repulsive Coulomb interactions \cite{Raghu2012}. A test of this idea would be to investigate multilayered films where one of the partners is non-superconducting and highly polarizable \cite{Berg2008}.

High T$_c$ superconductivity is reported to occur in very oxidized molybdenum cuprates. This system has received attention as the high T$_c$ may remain in a very overdoped region of the phase diagram, a region which has normally been considered to be simple Fermi liquids. 
The molybdenum substituted double chain cuprate Mo123 with the formula (Cu$_{0.75}$Mo$_{0.25}$Sr$_2$(Y)Cu$_2$O$_{7+\delta}$, is related to the well known YBa$_2$Cu$_3$O$_x$ (Y123) double chain cuprates in which Sr is substituted for Ba and in every other CuO chain 50\% of the Cu ions have been replaced by Mo. Three independent syntheses in Japan and Finland by Karpinnien, Yamauchi and coworkers over the past decade have reported consistently high T$_c$s. But they have not been repeated elsewhere perhaps because of the highly reactive oxidation at high temperatures, and pressures that are required.\cite{Karppinen2005,Grigoraviciute2006,Chmaissem2010,Marezio2013}.

\begin{figure}
\includegraphics[width=1\textwidth]{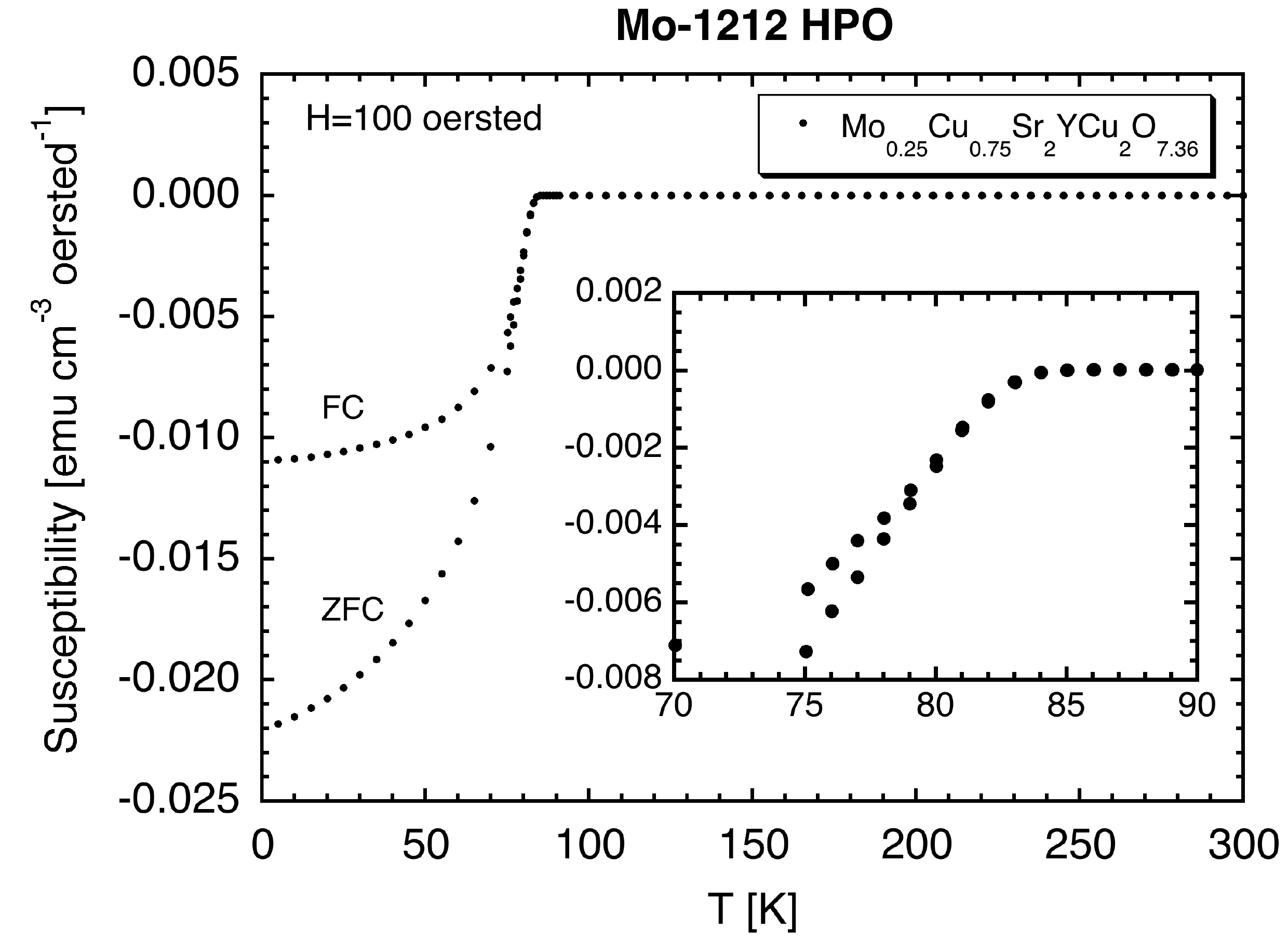}
\caption{\label{fig:Mo1212} Susceptibility versus temperature for the high pressure oxygenated (HPO) Mo-1212 cuprate. Field cooled (FC) and zero field cooled (ZFC) measurements show a diamagnetic transition beginning at $T_c = 86$ K. The inset shows a close-up of the transition. Figure reprinted with permission from \cite{Gauzzi2014}.}
\end{figure}

Recent measurements of the heat capacity and magnetism \cite{Gauzzi2014} show the superconductivity is a bulk effect, see Fig. \ref{fig:Mo1212}. Furthermore the refined neutron diffraction data confirm previous reports that upon the oxidation process superconductivity results \cite{Chmaissem2010}, and the apical oxygen distance is reduced to 2.15 $\AA$, much lower than in YBCO and other cuprates, and contrary to the positive correlation between T$_c$ and the apical oxide distance believed to obtain for cuprates with T$_c$s under dome \cite{Kuroki2011}. The increased positive charge on the highly oxidized CuO$_2$ layers gives a plausible explanation. A measurement of the superfluid density is needed to obtain direct evidence for a new pairing mechanism.\footnote{It is conceivable that only a fraction of the holes introduced by the oxidation become mobile and contribute to the superfluid density and the rest are localized. The T$_c$ is similar to optimally doped cuprates but the band structure must be different.}

A second possibly very overdoped cuprate is Sr$_2$CuO$_{4-x}$ that crystallizes in the same K$_2$NiO$_4$ structure as the (LaSr)$_2$CuO$_4$ family. It has been reported to be superconducting by Hiroi \cite{Hiroi1993, Han1994, Wang1995} and many studies since have reported values of $x$ that correspond to very overdoped superconductors, but uncertainties in $x$, the presence of superstructures with 4, 5, and incommensurate periods and possible contamination from oxidizing agents have clouded the issues. The recent work by Jin et al using SrO$_2$ as the oxidizing agent has eliminated the latter concern. \cite{Gao2009}

For more discussions of T$_c$s forming the dome, we refer the reader to other chapters in the Special Issue \cite{Chu2015}.

\section{Other Superconductors}

Superconductivity was discovered in sodium doped tungsten bronzes by Raub et al. \cite{Raub1964} and studied in detail for other alkali metals M$_x$WO$_3$ with M = K, Rb, Cs with T$_c$s ranging from 1-6 K \cite{Cadwell1981a,Skokan1979,Stanley1979a}. Remeika et al. found that T$_c$ increased with a decreasing Na, K, and Rb metal concentration until samples were no longer stable \cite{Remeika1967}. These studies suggest that epitaxial stabilization might allow further reduction of the alkali metal concentration and cause further increases in T$_c$. Wu et al. showed a thickness dependent superconductor-insulator transition \cite{Wu2014a} in potassium-doped K$_{0.33}$WO$_3$ thin films, and unveiled disorder enhanced Coulomb interactions to be present in the normal state. As we discuss below, we believe it worthwhile to investigate thin films of this system at lower alkali concentration.

The recent observation of large superconducting-like energy gap, which opens at temperatures close to the boiling point of liquid nitrogen, in monolayer FeSe films on SrTiO$_3$ (FeSe/STO) has generated tremendous interest \cite{Wang2012}. For more in depth reviews of the Fe-pnictide \cite{Kamihara2008} and Fe-chalcogenide \cite{Hsu2008a} superconductors, we refer readers to the respective reviews in this Special Issue. In the monolayer FeSe/STO, T$_c$s measured by transport have varied from 27 K \cite{Zhang2014} to 100 K \cite{Ge2014}, but the superconducting gap as measured by tunneling and ARPES is consistently seen to open around 50-65 K \cite{Liu2012,He2013,Tan2013}. Although there is still no consensus on the mechanism, it is clear that the interface between the unit cell of FeSe and SrTiO$_3$ plays a critical role. The STO substrate either strains FeSe via lattice mismatch giving rise to high T$_c$ (it is known that pressure raises bulk T$_c$ substantially \cite{Mizuguchi2008}) or transfers charges to FeSe. The high T$_c$s are found typically after careful annealing of the thin films, suggesting that effects on the interface due to selenium vacancies in the FeSe or oxygen vacancies in the STO must be carefully considered. There is certainly more to come from looking at T$_c$ in this system, as is discussed in the review on chalcogenides in this volume \cite{Chang2015}. 

\section{Intriguing USOs}

In this section we highlight a few cases of intriguing USOs, and we refer the reader to a more comprehensive review in this volume \cite{Kopelevich2015}. 

\subsection{Surface Doped Tungsten Bronzes}

Tantalizing reports of 90 K superconductivity in lightly surface doped Na$_{0.3}$WO$_3$ \cite{Reich1999,Reich2000a,Aliev2008a} and 120 K superconductivity in H$_x$WO$_3$ \cite{Reich2009a} require further investigation. Signatures of high T$_c$ were seen in resistance, magnetization and tunneling measurements. Electron spin resonance measurements of the Na-doped WO$_3$ gave evidence of weakly coupled high T$_c$ puddles of superconductivity \cite{Shengelaya1999}. These results have yet to be confirmed; part of the difficulty may lie in stabilizing the superconducting phase in bulk crystals but advanced thin film approaches may provide a route for stabilizing the high T$_c$ phase. We have laid the ground work by synthesizing superconducting K$_x$WO$_3$ films \cite{Wu2014a}, as well as stabilizing crystalline un-doped hexagonal WO$_3$ thin films \cite{Wu2014b} and employing ionic liquid gating studies to reach similar carrier concentrations, on the order of $10^{21} /$cm$^3$ carriers. In the K-doped samples we achieve T$_c$s $\sim$3-4 K, in contrast no T$_c$s were found with the ionic gating implying that to achieve superconductivity it is necessary to have ions, even in small quantities, in the hexagonal channels to induce superconductivity in this system; further investigations are warranted.

\subsection{Cu/CuO}

Copper monoxide (CuO) is an antiferromagnetic insulator that becomes the key layer in all the high T$_c$ cuprates. It was unexpected when Osipov et al. reported superconducting-like signals at temperatures as high as 400 K in Cu/CuO interfaces \cite{Osipov2001}. The samples consisted of elemental Cu thin films deposited on single crystal insulating CuO substrates. The interfaces were heated and quenched by several ampere current pulses of $\sim$10 ms duration. The samples that survived showed diamagnetic signals and very large critical currents well above room temperature. 

These results have not been reproduced so far as we know. The reported results could be due to inhomogeneities or bad measurements, though the authors state that both two probe and four probe measurements were consistent, providing credibility to the results. Ko et al. investigated Cu/CuO interfaces prepared under more controlled conditions \cite{Munakata2011}. Briefly, a CuO film of approximately 20 nm thickness was deposited on specially treated MgO substrate and followed by a 3nm layer of elemental Cu. The (111) direction of CuO was aligned with the (001) MgO. Details are given in the reference. No superconductivity was detected, however magnetoconductance measurements of the bilayer gave evidence for an unusual antiferromagnetic (AF) proximity effect. This nonlocal effect implies AF spin ordering in the Cu layer on top of the CuO and was anticipated theoretically \cite{Sherman2010}. Despite the disappointment in a null superconducting result, unusual physics was discovered.

\subsection{CuCl}
CuCl under pressure has given evidence for meta-stable superconductivity. In one report, transitions to nearly perfect diamagnetism around 100 K in bulk CuCl under pressure was shown to be stable for several hours \cite{Brandt1978}. In another report, evidence was reproducibly observed for a transient meta-stable diamagnetic signal and resistive transition at $\sim$180 K when warming CuCl crystals \cite{Chu1978,Geballe1979}. While these experimental results have yet to be reproduced and stabilized elsewhere, theoretical models have been proposed to explain the superconductivity \cite{Abrikosov1978,Rhim2007}. We believe it is worth to investigate CuCl based thin films and superlattice structures.


\section{Conclusion}
We have provided examples in which the T$_c$s themselves and their dependence upon external parameters have provided valuable insights of the underlying mechanisms. We have selected only a small portion of the extensive research that has been done, where further investigation is called for. We suggest that in some of these cases it will be possible to find more superconductors, with very high T$_c$ phases. The highly oxidized cuprates that have been discussed here are promising. Non-equilibrium phases obtained at interfaces and by quenching represent a huge phase space with opportunities for new discoveries. Negative U ions are effective pairing centers and could lead to very high T$_c$ provided the competing charge density wave order can be suppressed. Some of the unidentified T$_c$s, of which we have given a few promising examples, require further investigation. We conclude that searching for new and enhanced T$_c$'s remains a promising and never ending fertile frontier. 

\section*{Acknowledgments}
We would like to acknowledge helpful comments from Steve Kivelson, Sri Raghu, Ian Fisher, Mac Beasley, Massimo Marezio, and Jorge Hirsch before and during the preparation of this manuscript. We thank AFOSR under DoD MURI Grant FA9550-09-1-0583 for support.

\section*{References}

\bibliography{jphyscbib}

\end{document}